\documentstyle[12pt,prd,aps,epsfig,amssymb,preprint]{revtex}

\begin{document}
\title{{\bf A Systematic Study on Nonrelativistic Quarkonium Interaction }}
\author{Sameer M. Ikhdair\thanks{%
Electronic address (internet): sameer@neu.edu.tr} and \ Ramazan Sever\thanks{%
sever@metu.edu.tr}}
\address{$^{\ast }$Department of Electrical and Electronic Engineering, Near East\\
University, Nicosia, North Cyprus, Turkey\\
$^{\dagger }$Department of Physics, Middle East Technical University,
Ankara, Turkey.\\
PACS NUMBER(S): 12.39.Pn, 14.40-n, 12.39.Jh, 14.65.-q, 13.30.Gd
method, Quarkonia, Mesons}
\date{\today
}
\maketitle

\begin{abstract}
A recently proposed strictly phenomenological static quark-antiquark
potential belonging to the generality $V(r)=-Ar^{-\alpha }+\kappa r^{\beta
}+V_{0}~$is tested with heavy quarkonia in the context of the shifted large
N-expansion method. This nonrelativistic potential model fits the
spin-averaged mass spectra of the $c\overline{c},$ $b\overline{b}$ and $c%
\overline{b}$ quarkonia within a few ${\rm MeV}$\ and also the five
experimentally known leptonic decay widths of the $c\overline{c}$ and $b%
\overline{b}$ vector states. Further, we compute the hyperfine splittings of
the bottomonium spectrum as well as the fine and hyperfine splittings of the
charmonium spectrum. We give predictions for not yet observed $B_{c}$
splittings. The model is then used to predict the masses of the remaining
quarkonia and the leptonic decay widths of the two pseudoscalar $c\overline{b%
}$ states. Our results are compared with other models to gauge the
reliability of the predictions and point out differences.
\end{abstract}

\begin{center}
$\bigskip $
\end{center}

\begin{verbatim}

\end{verbatim}

\section{INTRODUCTION}

\noindent The charm-beauty ($B_{c})$ quarkonium states provide a
unique window into heavy quark dynamics. The properties of the
$B_{c}$ mesons are of special interest, since they are the only
quarkonia consisting of two heavy quarks with different flavours and
are also intermediate to the charmonium and bottomonium systems.
Additionally, because they carry flavour they cannot annihilate into
gluons so are more stable with widths less than a hundred ${\rm
keV.}$ Excited $B_{c}$ meson states lying below ${\rm BD}$ (and
${\rm BD}^{\ast }$ or ${\rm B}^{\ast }{\rm D}$) threshold can only
undergo hadronic (pionic) or radiative transitions to the ground
pseudoscalar state which then decays weakly. This results in a rich
spectroscopy of narrow radial and orbital excitations which are
considerably more stable than their charmonium and bottomonium
analogues. The discovery of the $B_{c}$ meson by the Collider
Detector at Fermilab (CDF) Collaboration [1]\ in $p\overline{p}$\
collisions at $\sqrt{s}=1.8$ $TeV$ with an observed pseudoscalar
mass $M_{B_{c}}({\rm 1S})=6.40\pm 0.39\pm 0.13$ ${\rm GeV}$ has
inspired new theoretical interest in the study of the $B_{c}$
spectroscopy in the framework of heavy quarkonium theory
[2,3,4,5,6,7,8,9,10,11,12,13,14,15].

On the other hand, for bottomonium system, the ALEPH collaboration in 2002
has searched for the pseudoscalar bottomonium meson, $\eta _{b},$ in
two-photon interactions at LEP2 with an integrated luminosity of $699$ pb$%
^{-1}$ collected at $e^{+}e^{-}$ centre-of mass energies from $181$ ${\rm GeV%
}$ to 209 ${\rm GeV.}$ One candidate event is found in the
six-charged-particle final state and none in the four-charged-particle final
state. The candidate $\eta _{b}({\rm 1S}$) ($\eta _{b}\rightarrow
K_{S}K^{-}\pi ^{+}\pi ^{-}\pi ^{+}$) has reconstructed invariant mass of $%
9.30\pm 0.02\pm 0.02$ ${\rm GeV}$ [16]${\rm .}$ Theoretical estimates (from
perturbative QCD and lattice nonrelativistic QCD) to the hyperfine mass
splitting for the ${\rm 1S}$ bottomonium state $\Delta _{{\rm HF}}({\rm 1S})$
are reported (cf. [16] and references therein).

Moreover, the inconsistent Crystal Ball measurement for the pseudoscalar
charmonium mass, $M_{\eta _{c}({\rm 2S})}=3594\pm 5$ ${\rm MeV}$ [17] has
been given for more than 20 years, before a new measurement of $3654\pm 6\pm
8$ ${\rm MeV}$ was available by Belle Collaboration in summer of 2002 [18]
for the exclusive $B\longrightarrow KK_{S}K^{-}\pi ^{+}$ decays. It is close
to the $\eta _{c}({\rm 2S})$ mass observed by the same group in the
experiment $e^{+}e^{-}\longrightarrow J/\psi \eta _{c}$ where $M_{\eta _{c}(%
{\rm 2S})}=3622\pm 12$ ${\rm MeV}$ was found [19]. It is giving rise to a
small hyperfine splitting for the ${\rm 2S}$ charmonium state, $\Delta _{%
{\rm HF}}({\rm 2S,}$exp)$=M_{\psi ({\rm 2S})}-M_{\eta _{c}({\rm 2S)}}=32\pm
14$ ${\rm MeV}$ [20]${\rm .}$

Very recently, several more new measurements have been appeared that support
the Belle value: BaBar measures $3630.8\pm 3.4\pm 1.0$ ${\rm MeV}$ [21] (an
earlier analysis resulted in $3632.2\pm 5.0\pm 1.8$ ${\rm MeV}$ [22]), CLEO
II gives $3642.7\pm 4.1\pm 4.0$ ${\rm MeV}$ [23] (CLEO III prelim: $%
3642.5\pm 3.6\pm ?$ ${\rm MeV})$ and a different Belle analysis yields $%
3630\pm 8$ ${\rm MeV}$ [24]. The $\Delta _{{\rm HF}}({\rm 2S})$ splitting
calculated from a naive average of the central values of all measurements
except Crystal Ball is $47$ ${\rm MeV}$). Further, the following mass values
have been obtained: $M_{J/\psi ({\rm 1S})}=3096.917\pm 0.010\pm 0.007$ ${\rm %
MeV}$ and $M_{\psi ({\rm 2S})}=3686.111\pm 0.025\pm 0.009$ ${\rm MeV.}$ The
relative measurement accuracy reached $4\times 10^{-6}$ for the $J/\psi (%
{\rm 1S}),$ $7\times 10^{-6}$ for the $\psi ({\rm 2S})$ and is approximately
3 times better than that of the previous precise experiments in [25] and
[26]. The new result for the mass difference is $M_{\psi ({\rm 2S}%
)}-M_{J/\psi ({\rm 1S})}=589.194\pm 0.027\pm 0.011$ ${\rm MeV}$ [2]${\rm .}$

Consequently, such observations of hyperfine splittings for the ${\rm 2S}$
charmonium and ${\rm 1S}$ bottomonium spectra have inspired new theoretical
interest in the study of the hyperfine splittings of the charmonium and
bottomonium states as well as their spectra [5,14,27,28,29]. Badalian and
Bakker [27] calculated the hyperfine splitting for the ${\rm 2S}$ charmonium
state, $\Delta _{{\rm HF}}({\rm 2S,}$theory)$=57\pm 8$ ${\rm MeV,}$ in their
recent work. Recksiegel and Sumino developed a new formalism [28] based on
improved perturbative QCD approach to compute the fine and hyperfine
splittings of charmonium and bottomonium [28,29].

As a result, one is able to test the validity of the conventional
phenomenological potential models where the core potential follow from
simple ansatzes by comparing our theoretical predictions of the spectrum of
heavy quarkonia in terms of agreement with the experimental data with
respect to the estimated uncertainties. Thus, these phenomenological
potential models are not a priori connected to a fundamental QCD parameters
[4,5,6,7,8]. In general, where the masses of bottomonium and charmonium
states have been measured, the experimental uncertainties are much smaller
than the theoretical uncertainties [20]. The only two exceptions are the
very poorly measured masses of $\eta _{b}({\rm 1S})$ and $\eta _{c}({\rm 2S}%
).$ With suitable potential model, very good agreement with the observed
spectra can be obtained for the charmonium and bottomonium states (e.g.,
[15]). These studies established the non-relativistic nature of the heavy
quarkonium systems and, in overall, a unified shape of the inter-quark
potential in the distance region $0.5\lesssim r\lesssim 5$ ${\rm GeV}^{-1}.$

In this work we extend our previous analysis of the shifted large-${\rm N}$
expansion technique (SLNET) developed for the Schr\"{o}dinger wave equation $%
\left[ 4,30,31,32\right] $ and then applied to semirelativistic and
relativistic wave equations [5,32,33] to reproduce the hyperfine splittings
of the $c\overline{c},$ $b\overline{b}$ and $c\overline{b}$ spectra using a
recently proposed phenomenological potential [15]. The calculations of the
bottomonium hyperfine splittings constitute predictions of the yet
unobserved states. The motivation of the present work is to give a detailed
analysis of mass spectra and decay widths for $c\overline{c},$ $b\overline{b}
$ and $B_{c}$ systems using a recently proposed phenomenological potential
form [15]. We also calculate the masses of the recently found new charmonium
$\eta _{c}({\rm 2S})$ and the searched bottomonium $\eta _{b}({\rm 1S})$
mesons together with the hyperfine splittings of their states.

The outline of this paper is as following: In Section II, we first review
briefly the analytic solution of the Schr\"{o}dinger equation for unequal
mass case $(m_{q_{i}}\neq m_{q_{j}})$. Section III is devoted for the
spin-averaged quarkonium masses of spin triplet states. The leptonic decay
widths are briefly reviewed in Section IV. Finally, Section V contains our
conclusions. \

\section{Schr\"{o}dinger Mass Spectrum}

\noindent We limit our discussion to the following generality of potentials
[5,15,34]:

\begin{equation}
V(r)=-Ar^{-\alpha }+\kappa r^{\beta }+V_{0},~
\end{equation}
where $A,$ $\kappa ,$ $\alpha $ and $\beta $ are nonnegative constants
whereas $V_{0}$ is taking any sign. These static quarkonium potentials are
monotone nondecreasing, and concave functions satisfying the condition
[4,5,30,32,33]

\begin{equation}
V^{\prime }(r)>0\text{ \ and }V^{\prime \prime }(r)\leq 0.
\end{equation}
At least ten potentials of this generality, but with various values of the
parameters, have been proposed in the literature. Cornell potential has $%
\alpha =\beta =1,$ Lichtenberg potential has $\alpha =\beta =0.75,$ Song-Lin
potential has $\alpha =\beta =0.5,$ and the logarithmic potential of Quigg
and Rosner corresponds to $\alpha =\beta \rightarrow 0,$ have been recently
studied (cf. e.g., [4,5,6,30,32,33] and references therein). The Song's
potential used in [34] has $\alpha =\beta =2/3.$ Potentials with $\alpha
\neq \beta $ have also been popular. Thus, Martin potential $\alpha =0,$ $%
\beta =0.1$ [4,5,6,30,32,33], while Grant, Rosner and Rynes [35] prefer $%
\alpha =0.045,$ $\beta =0.$ Heikkil\"{a}, T\"{o}rnquist and Ono [36] tried $%
\alpha =1,$ $\beta =2/3.$ More successful potential known in literature as
Indiana potential [37] and the Richardson potential [38]. Recently, Motyka
and Zalewski [15] have also explored the quality of fit in the region $0\leq
\alpha \leq 1.2,$ $0\leq \beta \leq 1.1$ of the $\alpha ,\beta $ plane. They
choose the region with coordinates $\alpha =1,\beta =0.5$.

In the present work, in order to get a good fit, we test the second
generality (1) with $\alpha =1,\beta =0.5.$ Therefore, the nonrelativistic
phenomenological potential used by Motyka and Zelawiski [15] for the $q_{i}%
\overline{q}_{i}$ and $q_{i}\overline{q}_{j}$ systems has the form

\begin{equation}
V(r)=-\frac{0.325250}{r}+0.70638\sqrt{r}-0.78891,
\end{equation}
where $V(r),$ $\sqrt{r}$ and $r^{-1}$ are all in units of $GeV.$
Consequently, as $r\rightarrow 0,$ this potential has the $r^{-1}$
dependence corresponding to one gluon exchange. The expected part of the
potential which is linear in $r$ is not seen. Probably the bottomonia are
too small to reach sufficiently far into the asymptotic region of linear
confinement. Perhaps a more flexible potential would exhibit the linear
part, but one may be observing an effect of the expected screening of the
interaction between the heavy quarks by the light sea quarks [15]. The
corresponding potential (3) is very reasonable (cf. e.g., [15]).

The quark massese are

\begin{equation}
m_{c}=1.3959\text{ }GeV,\text{ \ \ \ \ }m_{b}=4.8030\text{ }GeV.\text{ \ \ \
\ }
\end{equation}
For the $c\overline{b}$ quarkonium we use the reduced mass

\begin{equation}
\mu _{cb}=\frac{m_{c}m_{b}}{m_{c}+m_{b}}=1.0816\text{ }GeV.
\end{equation}
We follow Ref.[15] in choosing the centers of gravity of the triplets for
the practical reasons that the masses of the spin singlets for the $b%
\overline{b}$ quarkonium are not known or very poorly measured.

For two particles system, we shall consider the ${\rm N-}$dimensional space
Schr\"{o}dinger equation for any spherically symmetric central potential $%
V(r)$. If $\psi ({\bf r})$ denotes the Schr\"{o}dinger's wave function, a
separation of variables $\psi ({\bf r})=Y_{l,m}(\theta ,\phi
)u(r)/r^{(N-1)/2}$ gives the following radial Schr\"{o}dinger equation ($%
\hbar =c=1)$ [4,6,30,31,32]

\begin{equation}
\left\{ -\frac{1}{4\mu }\frac{d^{2}}{dr^{2}}+\frac{[\overline{k}-(1-a)][%
\overline{k}-(3-a)]}{16\mu r^{2}}+V(r)\right\} u(r)=E_{n,l}u(r),
\end{equation}
with $\mu =\frac{m_{q_{i}}m_{q_{j}}}{m_{q_{i}}+m_{q_{j}}}$ is the reduced
mass for the two quarkonium composite particles. Here, $E_{n,l}$ denotes the
Schr\"{o}dinger binding energy of meson, and $\overline{k}=N+2l-a,$ with $a$
representing a proper shift to be calculated later on and $l$ is the angular
quantum number. We follow the shifted $1/N$ or $1/\overline{k}$ expansion
method [4,5,30,31,32,33] by defining
\begin{equation}
V(x(r_{0}))\;=\stackrel{\infty }{%
\mathrel{\mathop{\sum }\limits_{m=0}}%
}\left( \frac{d^{m}V(r_{0})}{dr_{0}^{m}}\right) \frac{\left( r_{0}x\right)
^{m}}{m!Q}\overline{k}^{(4-m)/2},
\end{equation}
and also the energy eigenvalue expansion [4,5,30,32,33]

\begin{equation}
E_{n,l}\;=\stackrel{\infty }{%
\mathrel{\mathop{\sum }\limits_{m=0}}%
}\frac{\overline{k}^{(2-m)}}{Q}E_{m},
\end{equation}
where $x=\overline{k}^{1/2}(r/r_{0}-1)$ with $r_{0}$ is an arbitrary point
where the Taylor's expansions is being performed about and $Q$ is a scale
parameter to be set equal to $\overline{k}^{2}$ at the end of our
calculations. Following the approach presented by Ref.[4,30,32], we give the
necessary expressions for calculating the binding energies:

\begin{equation}
E_{0}=V(r_{0})+\frac{Q}{16\mu r_{0}^{2}},
\end{equation}
\begin{equation}
E_{1}=\frac{Q}{r_{0}^{2}}\left[ \left( n_{r}+\frac{1}{2}\right) \omega -%
\frac{(2-a)}{8\mu }\right] ,
\end{equation}
\begin{equation}
E_{2}=\frac{Q}{r_{0}^{2}}\left[ \frac{(1-a)(3-a)}{16\mu }+\alpha ^{(1)}%
\right] ,
\end{equation}
\begin{equation}
E_{3}=\frac{Q}{r_{0}^{2}}\alpha ^{(2)},
\end{equation}
where $\alpha ^{(1)}$ and $\alpha ^{(2)}$ are two useful expressions given
by Imbo {\it et al }[31] and also the scale parameter $Q$ is defined by the
relation

\begin{equation}
Q=8\mu r_{0}^{3}V^{\prime }(r_{0}).
\end{equation}
Thus, for the $N=3$ physical space, the Schr\"{o}dinger binding energy to
the third order is [4,30]

\begin{equation}
E_{n,l}=V(r_{0})+\frac{1}{2}r_{0}V^{\prime }(r_{0})+\frac{1}{r_{0}^{2}}\left[
\frac{(1-a)(3-a)}{16\mu }+\alpha ^{(1)}+\frac{\alpha ^{(2)}}{\overline{k}}%
+O\left( \frac{1}{\overline{k}^{2}}\right) \right] .
\end{equation}
where the shifting parameter, $a$, is defined by

\begin{equation}
a=2-(2n_{r}+1)\left[ 3+\frac{r_{0}V^{\prime \prime }(r_{0})}{V^{\prime
}(r_{0})}\right] ^{1/2},
\end{equation}
and the root, $r_{0},$ is being determined via

\begin{equation}
1+2l+(2n_{r}+1)\left[ 3+\frac{r_{0}V^{\prime \prime }(r_{0})}{V^{\prime
}(r_{0})}\right] ^{1/2}=\left[ 8\mu r_{0}^{3}V^{\prime }(r_{0})\right]
^{1/2},
\end{equation}
with $n_{r}=n-1$ is the radial quantum number and $n$ is the principal
quantum number. Once $r_{0}$ is found via equation (16), then the
Schr\"{o}dinger binding energy of the $q_{i}\overline{q}_{j}$ system in (14)
becomes relatively simple and straightforward. Hence, the bound state mass
of the $q_{i}\overline{q}_{j}$ system is written as

\begin{equation}
M(q_{i}\overline{q}_{j})_{nl}=m_{q_{i}}+m_{q_{j}}+2E_{n,l}.
\end{equation}
where $m_{q_{i}}$ and $m_{q_{j}}$ are the masses of the quark and antiquark,
respectively. The expansion parameter $1/N$ or $1/\overline{k}$ becomes
smaller as $l$ becomes larger since the parameter $\overline{k}$ is
proportional to $n$ which it appears in the denominator in higher-order
correction.

\section{Spin-averaged masses of spin triplet states}

Since the systems\ that we investigate in the present work are often
considered as nonrelativistic system, then our treatment is based upon
Schr\"{o}dinger equation with a Hamiltonian
\begin{equation}
H_{o}=-\frac{\triangledown ^{2}}{2\mu }+V(r)+V_{SS},
\end{equation}
where $V_{SS}$ is the spin-dependent term. The spin dependent correction to
the nonrelativistic Hamiltonian, which is responsible for the hyperfine
splitting of the mass levels, is generally used in the form [7,8,39,40,41,42]

\begin{equation}
V_{SS}\longrightarrow V_{{\rm HF}}=\frac{32\pi \alpha _{s}}{9m_{q_{i}}m_{%
\overline{q}_{j}}}({\bf s}_{1}.{\bf s}_{2}-\frac{1}{4})\delta ^{3}({\bf r}),
\end{equation}
adapted from the Breit-Fermi Hamiltonian. The number $\frac{1}{4}$
substituted from the product of the spins corresponds to the recent
assumption that the unperturbed nonrelativistic Hamiltonian gives the energy
of the triplet states. Since for the states with orbital angular momentum $%
L>0$ the wave function vanishes at the origin, the shift effect only the $%
{\rm S}$ states. Thus, the only first order effect of this perturbation is
to shift the $^{1}S_{0}$ states down in energy by

\begin{equation}
\Delta E_{{\rm HF}}=\frac{32\pi \alpha _{s}}{9m_{q_{i}}m_{\overline{q}_{j}}}%
\left| \psi (0)\right| ^{2},
\end{equation}
with the wave function at the origin is calculated by using the expectation
value of the potential derivative via [4,5,6,30,42,43]

\begin{equation}
\left| \psi (0)\right| ^{2}=\frac{\mu }{2\pi }\left\langle \frac{dV(r)}{dr}%
\right\rangle .
\end{equation}
In order to apply the last formula one needs the value of the wavefunction
at the origin-this is obtained by solving the Schr\"{o}dinger equation with
the nonrelativistic Hamiltonian and the coupling constant. In this approach,
the QCD strong coupling constant $\alpha _{s}(4\mu ^{2}),$on the
renormalization point $\mu ^{2}$ is not an independent parameter. It can be
connected (in the $\overline{MS\text{ }}$ renormalization scheme) through
the two-loop relation [10,15]
\begin{equation}
\alpha _{s}(\mu ^{2})=\frac{4\pi }{\beta _{0}}\frac{\left( \frac{2\mu }{%
\Lambda _{\overline{MS}}^{(n_{f})}}\right) ^{2}-1}{\ln \left[ \left( \frac{%
2\mu }{\Lambda _{\overline{MS}}^{(n_{f})}}\right) ^{2}\right] }
\end{equation}
where $\beta _{0}=11-\frac{2}{3}n_{f}.$ Like most authors (cf. e.g.
[4,5,6,7,8,15]), the strong coupling constant $\alpha _{s}(m_{c}^{2})$ is
fitted to the experimental charmonium hyperfine splitting number $\Delta _{%
{\rm HF}}(1{\rm S,}$exp$)=117\pm 2$ $MeV$ [4,20] yields

\begin{equation}
\alpha _{s}(m_{c}^{2})=0.254.
\end{equation}
Knowing the coupling at the scale $m_{c}^{2}$ we obtain the couplings at
other scales as follows. The number of flavours $(n_{f})$ is put equal to
three for $4\mu ^{2}\leq m_{c}^{2}$ (we are not interested in the region $%
4\mu ^{2}\leq m_{s}^{2}),$ equal to four for $m_{b}^{2}\geq 4\mu ^{2}\geq $ $%
m_{c}^{2}$ and equal to five for $4\mu ^{2}\geq $ $m_{b}^{2}$ (we are not
interested in the region $4\mu ^{2}\geq $ $m_{t}^{2}).$ Then the value of\ $%
\alpha _{s}(4\mu ^{2}=m_{c}^{2})$ from (22) is used to calculate $\Lambda _{%
\overline{MS}}^{(n_{f}=3)}$ and $\Lambda _{\overline{MS}}^{(n_{f}=4)}.$
Using the known value of $\Lambda _{\overline{MS}}^{(n_{f}=4)}$ and the
formula form (22) we find the value

\begin{equation}
\alpha _{s}(m_{b}^{2})=0.200,
\end{equation}
From this the value of $\Lambda _{\overline{MS}}^{(n_{f}=5)}$ is
found. Note that this supports our model since a different choice of
the Hamiltonian would in general lead to a different value of the
wave function at the origin and to a different determination of
$\alpha _{s}(m_{c}^{2})$ from the same hyperfine splitting. Then the
estimate of $\alpha _{s}(m_{Z}^{2})$
would, of course, be also different. For the hyperfine splitting of the $c%
\overline{b}$ quarkonium we use the coupling constant

\begin{equation}
\alpha _{s}(4\mu _{cb}^{2})=0.224,
\end{equation}
so that in each case the scale is twice the reduced mass of the
quark-antiquark system.

The calculated quarkonium masses together with hyperfine splittings are
given in Tables 1-3. The hyperfine mass splittings of $c\overline{c}$ and $b%
\overline{b}$ predicted by the potential model is listed with some other
models in Table 4. Therefore, as for the hyperfine splittings in Table 4,
all of the potential model calculations try to reproduce the old
experimental values, while the lattice calculations and perturbative QCD
favor the new values. No confirmed experimental data to check these
predictions are available yet. Let us note, however, that the unconfirmed
experimental splitting of the $2S(c\overline{c})$ level $-92/32$ $MeV$- is
much bigger/lower than expected from the potential models. In all cases,
where comparison with the other models are significantly smaller than the
splittings found by Eichten and Quigg [7] and similar to, but usually a
little smaller than, the splittings calculated by Gupta and Johnson [44].

One can also try to compare our results with more ambitious approaches. A
careful analysis in the framework of QCD sum rules [45] finds the hyperfine
splitting of the bottomonium $\Delta _{{\rm HF}}({\rm 1S,}$theory)$%
=63_{-51}^{+29}$ $MeV.$ The central value agrees to several ${\rm MeV}$ very
well with our expectation, but the uncertainty is too large to distinguish
between the potential models. A lattice calculation [9] gives the hyperfine
splitting $\Delta _{{\rm HF}}({\rm 1S,}$theory)$=60$ $MeV$ with a large
uncertainty. Again the central value is close to our model, but the
uncertainty is big enough to be consistent with all the potential models
quoted here.

\section{Leptonic Decay Widths}

The leading terms in the leptonic decay widths of the heavy quarkonia are
proportional to the squares of the wave functions at the origin. Therefore,
they are significant only for the $S$ states. For the $c\overline{c}$ and $b%
\overline{b}$ quarkonium systems, we shall consider the decays of the $%
n^{3}S_{1}$ (vector) states into pairs of charge conjugated charged leptons,
e.g. for definiteness into $e^{+}e^{-}$ pairs. For the $c\overline{b}$
quarkonium we consider the decays of the $n^{1}S_{0}$ (pseudoscalar) states
into $\tau \nu _{\tau }$ pairs. Since the probability of such decays
contains as a factor the square of the lepton mass, the decays into lighter
leptons are much less probable.

The decay widths of the vector $c\overline{c}$ and $b\overline{b}$
quarkonium systems into charged lepton pairs are usually calculated from the
QCD corrected Van Royen-Weiskopf formula [46]

\begin{equation}
\Gamma _{V\rightarrow \overline{l}l}=16\pi \alpha ^{2}e_{q}^{2}\frac{\left|
\psi (0)\right| ^{2}}{M_{V}^{2}}\left( 1-\frac{16\alpha _{s}(m_{q}^{2})}{%
3\pi }\right) .
\end{equation}
For vector mesons containing light quarks this formula leads to paradoxes
(cf. [47] and references therein). For quarkonia, however, the main problem
seems to be the QCD correction. Thus, in order to get quantitative
predictions it is necessary to include higher order corrections which are
not known. In order to estimate the missing terms we tried two simple forms.
Exponentialization of the first correction

\begin{equation}
C_{1}(\alpha _{s}(m_{q}^{2}))=\exp (-\frac{16\alpha _{s}(m_{q}^{2})}{3\pi }),
\end{equation}
and Padeization

\begin{equation}
C_{2}(\alpha _{s}(m_{q}^{2}))=\frac{1}{1+\frac{16\alpha _{s}(m_{q}^{2})}{%
3\pi }}.
\end{equation}
We use the average of these two estimates as our estimate of the QCD
correction factor extended to higher orders. The difference between $C_{1}$
and $C_{2}$ is our crude evaluation of the uncertainty of this estimate. The
resulting leptonic widths are collected in Table 5. Further, we have the
relation

\begin{equation}
\Gamma _{V\rightarrow \overline{l}l}=\frac{9}{8}\frac{4m_{q}^{2}}{M_{V}^{2}}%
\frac{\alpha ^{2}e_{q}^{2}}{\alpha _{s}(m_{q}^{2})}C_{av}\Delta E_{{\rm HF}},
\end{equation}
where $C_{av}$ is the averaged QCD correction factor. With our choice of
parameters this formula reduces to

\begin{equation}
\Gamma _{V\rightarrow \overline{l}l}=F(q)\frac{4m_{q}^{2}}{M_{V}^{2}}\Delta
E_{{\rm HF}},
\end{equation}
with $F(c)=7.07\times 10^{-5}$ and $F(b)=2.43\times 10^{-5}$.

The formula for the leptonic widths of the pseudoscalar $c\overline{b}$
quarkonium reads

\begin{equation}
\Gamma _{\tau \nu _{\tau }}=\frac{G^{2}}{8\pi }f_{B_{c}}^{2}\left|
V_{cb}\right| ^{2}M_{B_{c}}m_{\tau }^{2}\left( 1-\frac{m_{\tau }^{2}}{%
M_{B_{c}}^{2}}\right) ^{2},
\end{equation}
where $G$ is the Fermi constant, $V_{cb}\approx 0.04$ is the element of the
Cabibbo-Kobayashi-Masakawa matrix and the decay constant $f_{B_{c}}$ is
given by the formula (cf. e.g. [48])

\begin{equation}
f_{B_{c}}^{2}=\frac{12\left| \psi (0)\right| ^{2}}{M_{B_{c}}^{2}}\overline{C}%
^{2}(\alpha _{s}),
\end{equation}
where $\overline{C}(\alpha _{s})$ is QCD correction factor. Formally this
decay constant is defined in terms of the element of the axial weak current

\begin{equation}
\left\langle 0\right| A_{\mu }(0)\left| B_{c}(q)\right\rangle
=if_{B_{c}}V_{cb}q_{\mu }.
\end{equation}
The QCD correction factor is [48]

\begin{equation}
\overline{C}(\alpha _{s})=1-\frac{\alpha _{s}(\mu _{cb}^{2})}{\pi }\left[ 2-%
\frac{m_{b}-m_{c}}{m_{b}+m_{c}}\ln \frac{m_{b}}{m_{c}}\right] .
\end{equation}
With our parameters $\overline{C}(\alpha _{s})\approx 0.905$ and since this
is rather close to unity, we use it without trying to estimate the higher
order terms.

Substituting the numbers one finds the decay widths given in Table 5. The
corresponding decay constants for the ground state and for the first excited
${\rm S-}$state of the $c\overline{b}$ quarkonium are found to be $%
f_{B_{c}}=492$ ${\rm MeV}$ and $f_{B_{c}}=338$ ${\rm MeV}$ (cf. e.g., [6].

Let us note the convenient relation

\begin{equation}
f_{B_{c}}^{2}=\frac{27\mu _{cb}}{8\pi \alpha _{s}(4\mu _{cb}^{2})}\frac{%
m_{b}+m_{c}}{M_{B_{c}}}\overline{C}^{2}(\alpha _{s})\Delta E_{{\rm HF}},
\end{equation}
which for our values of the parameters yields

\begin{equation}
f_{B_{c}}=65.2\sqrt{\frac{6199}{M_{B_{c}}}}\sqrt{\Delta E_{{\rm HF}}},
\end{equation}
where all the parameters are in suitable powers of ${\rm MeV}.$\ \ \ \ \ \ \
\ \ \ \ \ \ \ \ \ \ \ \ \ \ \ \ \ \ \ \ \ \

\section{\ \ Conclusions\ \ }

Figure1 shows the behavior of the present potential form in comparison with
some other potential forms considered in [4]. From this figure it is noticed
that, all potential forms are nearly the same at large $r,$ but the behavior
seem different at very small $r$ ($r<0.5$ ${\rm GeV}^{-1},$ about $0.1$ $%
fm), $ due to the variation of the one-gluon exchange term in each potential
model. Further, the present potential model containing six free parameters:
the three parameters in the strictly nonrelativistic phenomenological
potential (3), the masses of the $c$ and $b$ quarks (4) and the strong
coupling constant at the $m_{c}$ scale (22). This model is applicable to all
heavy quarkonia below their strong decay thresholds.

Consequently, we obtain the $c\overline{c},b\overline{b}$ and $c\overline{b}%
(b\overline{c})$ quarkonium mass spectra and also their leptonic decay
widths in good agreement with up-to-date experimental findings. We also give
prediction for the $c\overline{b}$ quarkonium masses and also for the
leptonic widths of the pseudoscalar $c\overline{b}$ quarkonium. Our model
predicts similar hyperfine splitting for the ${\rm 1S}$ bottomonium and $%
{\rm 2S}$ charmonium as in the other potential models [7,15], lattice
[49,50] and perturbation QCD [28,29,51]. Further, the fine splitting in
charmonium is found to be $M_{\psi ({\rm 1S})}-M_{J/\psi ({\rm 1S})}=597$ $%
MeV.$

Finally, in general, the potential models reproduce the experimental values
much better. This feature would be understandable, since the potential
models contain much more input parameters than the lattice or perturbative
QCD models.

\acknowledgments This research was partially supported by the
Scientific and Technical Research Council of Turkey. The authors
wish to thank S. Recksiegel for information on recent QCD-based
formalism. Sameer M. Ikhdair would like to thank his wife, Oyoun,
and his son, Musbah, for their assistance, patience and love.

\newpage
\begin{figure}[tbp]
\caption{The behavior of the different potential models versus $r.$}
\label{Figure1}
\end{figure}

\baselineskip= 2\baselineskip

\newpage

\begin{table}[tbp]
\caption{Schr\"{o}dinger bound-state mass spectrum of $c\overline{c}$
quarkonium (in $MeV)$. $\Delta X$ denotes the difference between the mass of
particle $X$ and the centre of gravity of the spin triplet part of the
multiplet, where $X$ belongs.}
\label{Table 1}
\begin{tabular}{llllll}
State & EQ [7] & GJ [44] & MZ [15] & This work & Expt.[20] \\
\tableline$1^{3}S_{1}$ & 3097 & 3097 & 3097 & 3097 & 3096.87$\pm 0.04$ \\
$\Delta 1^{1}S_{0}$ & -117 & -117 & -117 & -117 & -117 \\
$1P$(c.o.g) & 3492 & 3526 & 3521 & 3521 & 3525.3$\pm 0.2$ \\
$2^{3}S_{1}$ & 3686 & 3685 & 3690 & 3694 & 3685.96$\pm 0.09$ \\
$\Delta 2^{1}S_{0}$ & -78 & -68 & -72 & -65.9 & -92/-32 \\
$1D$(c.o.g) &  &  &  & 3806 & ($1^{3}D_{1}$ state) 3769.9$\pm 25$ \\
$2P$(c.o.g) &  &  &  & 3944 &  \\
$3^{3}S_{1}$ &  &  &  & 4078 & 4040$\pm 10$ \\
$2D$(c.o.g) &  &  &  & 4150 & 4159$\pm 20?$ \\
$3P$(c.o.g) &  &  &  & 4265 &  \\
$4^{3}S_{1}$ &  &  &  & 44377 & 4415$\pm 6$ \\
$3D$(c.o.g) &  &  &  & 4430 &  \\
$5^{3}S_{1}$ &  &  &  & 4628 &
\end{tabular}
\end{table}

\bigskip

\begin{table}[tbp]
\caption{Schr\"{o}dinger bound-state mass spectrum of $b\overline{b}$
quarkonium (in $MeV)$. $\Delta X$ denotes the difference between the mass of
particle $X$ and the centre of gravity of the spin triplet part of the
multiplet, where $X$ belongs.}
\label{Table 2}
\begin{tabular}{llllll}
State & EQ [7] & KR [11] & MZ [15] & This work & Expt. [20] \\
\tableline$1^{3}S_{1}$ & 9464 &  & 9460 & 9460 & 9460 \\
$\Delta 1^{1}S_{0}$ & -87 &  & -56.7 & -57.9 & (160) \\
$1P$(c.o.g) & 9873 & 9903 & 9900 & 9900 & 9900 \\
$2^{3}S_{1}$ & 10007 &  & 10023 & 10031 & 10023 \\
$\Delta 2^{1}S_{0}$ & -44 &  & -28 & -23.2 &  \\
$1D$(c.o.g) & 10127 & 10156 & 10155 & 10155 & . \\
$2P$(c.o.g) & 10231 & 10259 & 10260 & 10261 & 10260 \\
$3^{3}S_{1}$ & 10339 & - & 10355 & 10364 & 10355 \\
$2D$(c.o.g) & - & 10441 & 10438 & 10438 &  \\
$3P$(c.o.g) & - & 10520 & 10525 & 10527 &  \\
$4^{3}S_{1}$ &  &  &  & 10614 &  \\
$3D$(c.o.g) &  &  &  & 10666 &  \\
$5^{3}S_{1}$ &  &  &  & 10820 &
\end{tabular}
\end{table}

\bigskip \bigskip
\begin{table}[tbp]
\caption{Schr\"{o}dinger bound-state mass spectrum of $c\overline{b}$ $(b%
\overline{c})$ quarkonium in $MeV$. $\Delta X$ denotes the difference
between the mass of particle $X$ and the centre of gravity of the spin
triplet part of the multiplet, where $X$ belongs.}
\label{Table 3}
\begin{tabular}{llllllll}
State & MZ [15] & CK [52] & EQ [7] & R [53] & G [12] & GJ [44] & This work
\\
\tableline$1^{3}S_{1}$ & 6349 & 6355 & 6337 & 6320 & 6317 & 6308 & 6349 \\
$\Delta 1^{1}S_{0}$ & -58 & -45 & -73 & -65 & -64 & -41 & -58.2 \\
$1P$(c.o.g) & 6769 & 6764 & 6736 & 6753 & 6728 & 6753 & 6769 \\
$2^{3}S_{1}$ & 6921 & 6917 & 6899 & 6900 & 6902 & 6886 & 6926 \\
$\Delta 2^{1}S_{0}$ & -33 & -27 & -43 &  & -35 & -33 & -30 \\
$1D$(c.o.g) &  &  & - &  &  &  & 7040 \\
$2P$(c.o.g) & 7165 & 7160 & 7160 &  & 7122 &  & 7165 \\
$3^{3}S_{1}$ &  &  &  &  &  &  & 7288 \\
$2D$(c.o.g) &  &  &  &  &  &  & 7359 \\
$3P$(c.o.g) &  &  &  &  &  &  & 7464 \\
$4^{3}S_{1}$ &  &  &  &  &  &  & 7567 \\
$3D$(c.o.g) &  &  &  &  &  &  & 7619 \\
$5^{3}S_{1}$ &  &  &  &  &  &  & 7800
\end{tabular}
\end{table}

\bigskip

\mediumtext

\bigskip \bigskip \bigskip
\begin{table}[tbp]
\caption{Level splittings in charmonium and bottomonium (in $MeV)$.}
\label{table1 4}
\begin{tabular}{llllllllll}
Level splitting\tablenotetext[1]{Potential model.} & Expt. & [7]%
\tablenotemark[1] & [15]\tablenotemark[1] & [14]\tablenotemark[1] & [54]%
\tablenotemark[1] & [49]\tablenotemark[2] & [50]\tablenotemark[2] & [29]%
\tablenotemark[3] & This work \\
\tableline$\Delta _{HF}^{(c\overline{c})}(2S)=M_{\psi (2S)}-M_{\eta
_{c}(2S)} $ & 92/32 & 78 & 72 & 98 & 92 & 43 & - & 38 & 66 \\
$\Delta _{HF}^{(b\overline{b})}(1S)=M_{\Upsilon (1S)}-M_{\eta _{b}(1S)}$%
\tablenotetext[2]{Lattice.} & (160) & 87 & 57 & 60 & 45 & - & 51 & 44 & 58
\\
$\Delta _{HF}^{(b\overline{b})}(2S)=M_{\Upsilon (2S)}-M_{\eta _{b}(2S)}$%
\tablenotetext[3]{Perturbative QCD.} & - & 44 & 28 & 30 & 28 & - & - & 21 &
23
\end{tabular}
\end{table}

\bigskip

\begin{table}[tbp]
\caption{Leptonic widths (in $KeV)$.}
\label{table1 5}
\begin{tabular}{lllll}
State & EQ [7] & MZ [15] & This work & Expt. \\
\tableline$1^{3}S_{1}(c\overline{c})$ & 8 & 4.5$\pm 0.5$ & 6.72$\pm 0.49$ &
5.3$\pm 0.4$ \\
$2^{3}S_{1}(c\overline{c})$ & 3.7 & 1.9$\pm 0.2$ & 2.66$\pm 0.19$ & 2.1$\pm
0.2$ \\
$1^{3}S_{1}(b\overline{b})$ & 1.7 & 1.36$\pm 0.07$ & 1.45$\pm 0.07$ & 1.32$%
\pm 0.05$ \\
$2^{3}S_{1}(b\overline{b})$ & 0.8 & 0.59$\pm 0.03$ & 0.52$\pm 0.02$ & 0.52$%
\pm 0.03$ \\
$3^{3}S_{1}(b\overline{b})$ & 0.6 & 0.40$\pm 0.02$ & 0.35$\pm 0.02$ & 0.48$%
\pm 0.08$ \\
$1^{1}S_{0}(c\overline{b})$ & 4$\times 10^{-8}$ & 2.8$\times 10^{-8}$ & 3.58$%
\times 10^{-8}$ & - \\
$2^{1}S_{0}(c\overline{b})$ & - & 1.6$\times 10^{-8}$ & 1.89$\times 10^{-8}$
& -
\end{tabular}
\end{table}
\newpage

\begin{figure}
\begin{center}
\epsfig{file=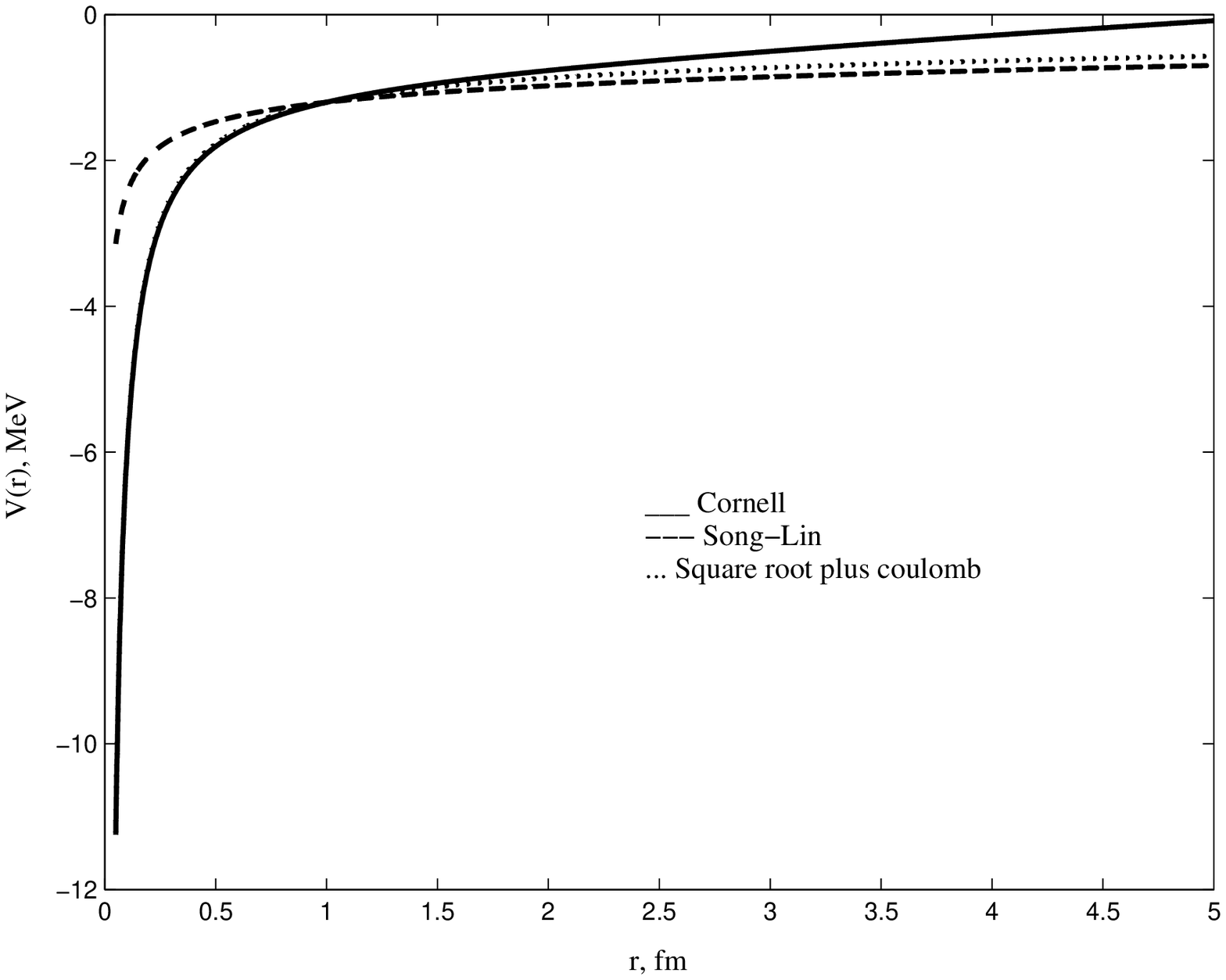,width=12cm,angle=0}
\end{center}
\end{figure}

\end{document}